\documentclass[letter]{aa}  

\usepackage{amsmath, amssymb}

\usepackage{txfonts}

\usepackage{graphicx}
\usepackage{graphics, epsfig, epsf}
\usepackage{grffile} 
\usepackage{longtable}
\usepackage{threeparttable, booktabs}
\usepackage{array}
\usepackage{multirow}
\usepackage{lscape}
\usepackage{rotating}
\usepackage{float}
\usepackage{placeins}

\usepackage{xcolor}
\usepackage{color}
\usepackage{soul}
\usepackage[normalem]{ulem}
\usepackage{textcomp}

\usepackage{placeins}
\usepackage{fancyhdr}

\usepackage{enumerate}

\usepackage{url}
\usepackage{wrapfig}
\usepackage{setspace}
\usepackage{appendix}
\usepackage{tikz}

\usepackage{breqn}
\usepackage{lipsum}

\def\apjl{\rm ApJL}
\def\apjs{\rm ApJS}

\def\aap{\rm A\&A}

\usepackage[colorlinks=true,linkcolor=blue,citecolor=blue]{hyperref}

   \title{A one-parameter two-zone leptonic model for the blazar sequence}

   \author{S. Boula\inst{1}\orcid{0000-0001-7905-6928}, A. Mastichiadis\inst{2}\orcid{0000-0001-5217-4801} \and D. Kazanas\inst{3}\orcid{0000-0002-7435-7809}
   }
   
   \institute{
INAF -- Osservatorio Astronomico di Brera, Via E. Bianchi 46, I-23807 Merate, Italy\\
\email{styliani.boula@inaf.it}
\and
Department of Physics, National and Kapodistrian University of Athens, University Campus Zografos, GR 15783, Athens, Greece 
\and NASA Goddard Space Flight Center, Greenbelt, MD, United States
             }

   \date{Received October 08, 2025; accepted December 12, 2025}

\begin{document} 
  \abstract
   {Blazars, a subclass of radio-loud active galactic nuclei with relativistic jets aligned close to our line of sight, emit highly variable non-thermal radiation across the electromagnetic spectrum. The physical origin of their emission and the blazar sequence remain open questions. We present a self-consistent two-zone leptonic model in which relativistic electrons accelerate in a compact region, losing energy via synchrotron and inverse Compton processes, and escape into a larger zone permeated by an external photon field associated with magnetohydrodynamic winds from the accretion disk. By varying only the mass accretion rate onto the central black hole, the model naturally reproduces the blazar sequence, including Compton Dominance, $\gamma$-ray spectral indices, and the positions of synchrotron and inverse Compton peaks, while variations in secondary parameters account for the observed spread in the data. Flat Spectrum Radio Quasars exhibit strong external Compton emission from the extended zone, whereas BL Lac objects are dominated by synchrotron and synchrotron self-Compton emission from the compact acceleration region. This framework highlights the key role of accretion rate and spatially structured emission zones in shaping blazar spectra and provides a unified interpretation of their diverse phenomenology.}


   \authorrunning{Boula et al.}
   \maketitle
%
\section{Introduction}
Blazars, comprising flat-spectrum radio quasars (FSRQs) and BL Lac objects, are among the most extreme active galactic nuclei (AGN), with relativistic jets closely aligned to our line of sight, \cite{Blandford19}. Their broadband emission is non-thermal and strongly Doppler-boosted by the bulk relativistic motion of the jet plasma. The spectral energy distribution (SED) of blazars displays two broad components: a low-energy hump, from radio to optical–UV (and occasionally X-rays), attributed to synchrotron radiation from relativistic electrons; and a high-energy hump, from X-rays to GeV–TeV $\gamma$-rays, generally interpreted as inverse Compton (IC) scattering of either the synchrotron photons themselves (Synchrotron Self-Compton, SSC) or photons from external fields (External Compton, EC) (e.g., \citep{dermer92, sikora94, tavecchio98, BM22}). Variability timescales, combined with relativistic beaming and radiative cooling constraints, provide important clues about the size and location of the emitting regions. Additionally, optical and multi-wavelength polarization measurements reveal the magnetic field structure and degree of ordering in the emitting regions, providing complementary constraints on particle acceleration and emission site geometry \citep{Liodakis2022}.

The non-thermal particle populations responsible for these SED components must be continuously replenished by in-situ acceleration processes within the jet, (for a review \citep{M2020}). Several such mechanisms have been proposed and investigated in the literature: diffusive shock acceleration at relativistic shocks \citep{Blandford1978,Kirk2000,Shirin25}; magnetic reconnection in highly magnetized plasma (for a review \citep{sironi2025}); shear (or shear-driven) acceleration operating in velocity-gradient layers of the flow(for a review \citep{rieger2019}); and stochastic (second-order Fermi) acceleration in turbulent regions, e.g. ,\cite{2006KGMT}, or combinations of them, e.g., \cite{tavecchio21}.
 The efficiency and relative importance of these processes depend on local magnetization, flow Lorentz factor and shear, turbulence amplitude and spectrum, and dissipation physics, and may vary across sources and along the jet \citep{tavecchio21}. 

A systematic trend in the SED properties—the so-called blazar sequence by \cite{Foss98} and the updated one by \cite{Ghis17}—shows that as bolometric luminosity $L_{\rm bol}$ increases, the synchrotron peak frequency $\nu_{\rm pk}^{\rm syn}$ decreases, while the Compton dominance (CD; the IC-to-synchrotron luminosity ratio) and $\gamma$-ray spectral slope both increase, \citep{Finke13}. The original sequence, based on a sample of 132 objects with only 33 $\gamma$-ray detections, has been refined with \emph{Fermi}-LAT observations, expanding the $\gamma$-ray–detected blazar sample to over 5000 sources \citep{4th} and suggesting that the sequence may ultimately reflect fundamental AGN parameters.

The optically thin nature of blazar GeV emission places the $\gamma$-ray production site far from the immediate vicinity of the black hole (BH), possibly at distances up to $\sim 10^6 R_S$ ($\sim$10 pc), for $M_{\rm BH} \sim 10^8 M_\odot$, where $R_S$ the Schwarzschild radius \citep{Marscher10}. On these scales, the dusty molecular tori invoked in AGN unification \citep{AntonMill} may interact significantly with the jet. A compelling picture identifies these structures as magnetohydrodynamic (MHD) accretion-disk winds \citep{KK94,CL94}, launched from a few $R_S$ out to the BH sphere of influence ($\sim 10^6 R_S$). Observations of warm absorbers—blue-shifted absorption features—successfully modeled as photoionized MHD winds extending to comparable distances \citep{Behar09,FKCB} support this interpretation. Similar winds inferred in stellar-mass black hole systems \citep{FKCB17} suggest a scale-invariant phenomenon.

The physical properties of these MHD winds depend primarily on a small set of parameters and, most notably, on the mass accretion rate. Previous work showed that varying only this parameter can, in principle, reproduce the main trends of the blazar sequence \citep{BKM19,20BMK}. 
To explain the observed multiwavelength spectrum, we adopt a two-zone framework.
In the compact acceleration region considered here, particle energization can occur, for instance, through diffusive shock acceleration, after which electrons escape into a larger-scale cooling zone where 
radiative losses dominate.
Building on this idea, we develop a two-zone emission model in which electrons are accelerated near the central engine and subsequently escape into a cooling zone, where interactions with the ambient photon field associated with the MHD wind can dominate the high-energy emission.
 
 This Letter is structured as follows: in Sect.~\ref{sec:2} we describe the emission model and its scalings with mass accretion rate; in Sect.~\ref{sec:3} we present the two-zone emission framework; and in Sect.~\ref{sec:4} we summarize our conclusions.

\section{Scaling of particle emission and acceleration with mass accretion rate}\label{sec:2}
In our previous work \citep{BKM19}, we modeled blazar emission 
with relativistic electrons radiating via synchrotron and Compton processes, 
the latter involving seed photons originating from the MHD wind. By varying only the mass accretion rate, the model reproduced observed correlations between the synchrotron peak, the gamma-ray index, and the Compton dominance. A natural next step is to include explicit particle acceleration physics, which can arise from shocks, turbulence, or shear flows \citep[e.g.,][]{KRM98,RiegerMannheim2002}.

To capture these effects, we introduce a characteristic acceleration timescale $t_{\rm acc}$, parameterized to depend on $\dot{m}$. For first-order Fermi acceleration at relativistic shocks, the timescale can be written 
 $t_{\rm acc,FI} \gtrsim \left(\frac{c}{u_s}\right)^2 \frac{\lambda}{c} \simeq \frac{r_g c}{u_s^2}$,
with $u_s$ the shock speed in the comoving frame, $\lambda$ the particle mean free path, and $r_g=\gamma m c^2/(eB)$ the gyroradius. 

The spectral shape of blazars depends on the ratio between magnetic and photon energy densities. Following \citet{BKM19}, we assume partial equipartition between magnetic and accretion power, with accretion luminosity $P_{\rm acc} = \dot{m}\,{\cal M}\,L_{\rm Edd}$, where $L_{\rm Edd}=1.28\times10^{38}~\rm erg\,s^{-1}$, $\dot{m}$ is the Eddington-normalized accretion rate, and ${\cal M}=M_{\rm BH}/M_\odot$. The magnetic energy density at distance $z$ then scales as $U_{\rm B} \propto \eta_{\rm b}\,\dot{m}\,{\cal M}^{-1}$, with $\eta_{\rm b}$ a proportionality constant.

We assume that the external photon field originates from accretion-disk photons scattered by the MHD wind. The scattering region extends from a radial distance $R_1$ to $R_2$ along the wind, with electron density $n(r) = n_0 (r/R_s)^{-1}$ following our MHD wind prescription \citep{CL94}. The energy density of scattered photons in the jet comoving frame is $U_{\rm ext} = \Gamma^2 U_{\rm sc}$, where 
$U_{\rm sc} \simeq L_{\rm disk} \tau_{\rm T}/(4\pi R_2^2 c)$ and the wind Thomson depth is $\tau_{\rm T} = n_0 \sigma_{\rm T} R_s \ln(R_2/R_1)$, where $n_0 \propto \dot{m}$. 
Here $L_{\rm disk} \propto \epsilon \dot{m}^\alpha {\cal M} L_{\rm Edd}$, with $\alpha=1$ for radiatively efficient disks (FSRQs) and $\alpha=2$ for ADAF-like disks (BL Lac obects), and $\epsilon$ is the radiative efficiency. 
Therefore, the photon field energy density scales as $U_{\rm ext} \propto \Gamma^2 \epsilon \dot{m}^{\alpha+1} {\cal M}^{-1}$.

The emission is assumed to arise from a stationary spherical zone of radius $R_{\rm b}$, located at some distance $z$ along the jet. The blob moves with bulk Lorentz factor $\Gamma$ at angle $\theta$ to the line of sight, giving Doppler factor $\delta=[\Gamma(1-\beta\cos\theta]^{-1}$. Particles are accelerated with a timescale $t_{\rm acc} \propto \gamma \, A_{\rm acc}(\dot{m})$. In the framework of diffusive (Fermi~I) shock acceleration, the acceleration timescale 
depends on the particle gyroradius and the shock speed, and can be indirectly linked to the accretion rate through the magnetic field strength, since $B \propto \dot{m}^{1/2}$ in 
standard jet–disk coupling scenarios. 
We initially tested this physical dependence, but found that it could not reproduce 
the observed trend of softer spectral peaks in high-accretion-rate sources. 
To investigate this effect without introducing additional free parameters, we therefore adopted a phenomenological scaling in which the acceleration timescale increases linearly with the accretion rate, $t_{\mathrm{acc}} \propto \dot{m}$. This prescription effectively captures the possibility that more powerful (high $\dot{m}$) systems may experience slower particle acceleration, for example due to reduced turbulence or more complex shock geometries.

\begin{figure}[!htbp]
\centering
\includegraphics[width=0.7\linewidth]{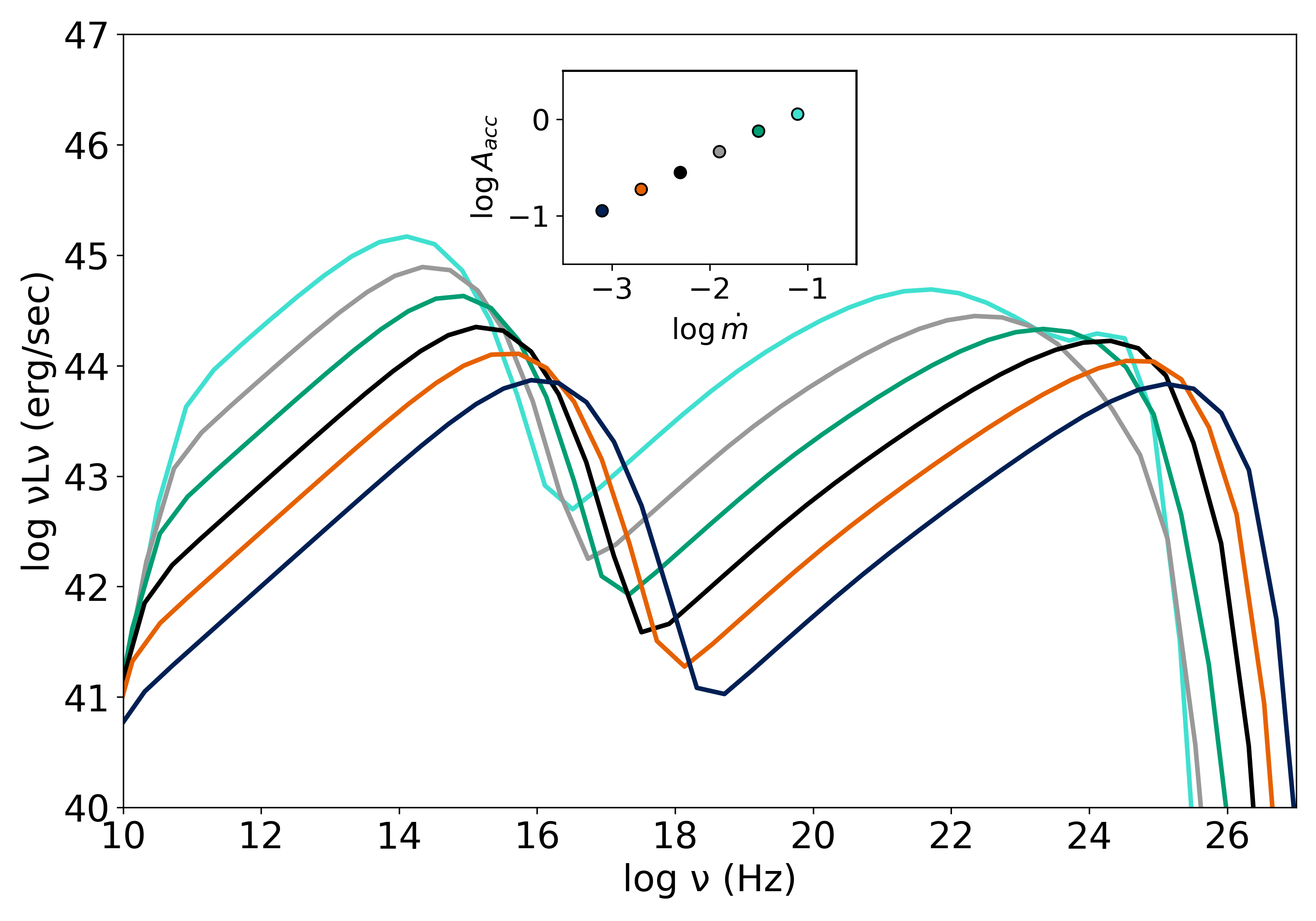}
\caption{
{Bottom:} Calculated Spectral Energy Distributions of BL Lac objects for various values of the normalized accretion rate $\dot{m}$.  
{Inset (top center):} Dependence of the acceleration timescale $t_{\rm acc}$ on $\dot{m}$, showing the nearly linear scaling required to explain BL Lac sources. Colored points correspond to the SED curves, indicating that higher $\dot{m}$ produces more luminous sources.
}
\label{fig:1a}
\end{figure}

The electron distribution in the acceleration region is governed by the kinetic equation:
\begin{equation}\label{eq:kinI}
\frac{\partial n_{e}(\gamma,t)}{\partial t} + \frac{n_{e}(\gamma,t)}{t_{\rm esc}} =
Q_{\rm inj} + \mathcal{L}_{\rm syn}(\gamma,t) + \mathcal{L}_{\rm ICS}(\gamma,t) +
\frac{\partial}{\partial \gamma}\!\left[\frac{\gamma\, n_e(\gamma,t)}{t_{\rm acc}}\right],
\end{equation}
where $n_e(\gamma,t)$ is the differential distribution, $t_{\rm esc}$ the escape timescale, $Q_{\rm inj}$ the injection term, and $\mathcal{L}_{\rm syn}$ and $\mathcal{L}_{\rm ICS}$ the synchrotron and inverse-Compton loss terms \citep{MK95}. Solving the coupled electron–photon equations for $\dot{m}\leq0.1$, we find that reproducing the blazar sequense requires $t_{\rm acc}$ to scale nearly linearly with $\dot{m}$ (Fig.~\ref{fig:1a}, inset panel).

This section establishes a one-zone leptonic framework where both the acceleration timescale and the external photon field depend on the mass accretion rate. In the next section, we extend this approach by embedding the acceleration region in a larger cooling zone, moving from a one-zone description to a two-zone emission model.

\section{Particle escape: A two-zone emission model}\label{sec:3}
\begin{figure}[!h]
    \centering
    \includegraphics[width=0.45\linewidth]{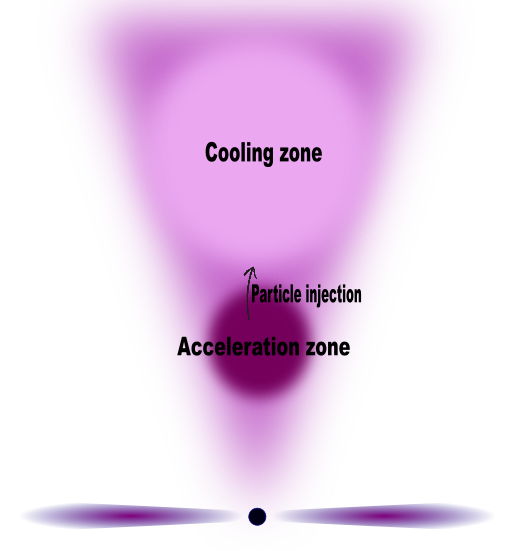}
    \caption{Sketch of the two-zone model. Particles accelerate and radiate in the acceleration zone (dark blue). A fraction escapes into the cooling zone (light blue), where it continues to radiate.}
    \label{fig:model_2zone}
\end{figure}

In order to take into account that the high energy electrons escaping the acceleration zone still suffer ics losses from the scattered disk photons on the wind, we extend the one-zone model presented in the previous secion onto a two-zone scenario.Thus, we consider that particles are accelerated in a compact region near the central engine (acceleration zone, zone I), then escape into a larger volume further along the jet (cooling zone, zone II), where they continue to radiate via synchrotron and inverse Compton processes (Fig.~\ref{fig:model_2zone}).

The kinetic equation for electrons in the acceleration zone is described by Eq. \ref{eq:kinI}.
In the cooling zone, the kinetic equation is
\begin{equation}
    \frac{\partial n_{e_{II}}(\gamma,t)}{\partial t} 
    + \frac{n_{e_{II}}(\gamma,t)}{t_{{\rm esc}_{II}}} 
    = Q_{\rm inj} + \mathcal{L}_{\rm loss}(\gamma,t),
\end{equation}
with $t_{{\rm esc}_{II}} = R_{II}/c$ set by the light-crossing time of the cooling zone, and $Q_{\rm inj} = \frac{n_{e_I}}{t_{{\rm esc}_I}}$ representing electrons escaping from zone I and injected into zone II. The acceleration and cooling regions in our model differ in size by about one order of magnitude 
($R_{II} \simeq 10 R_{I}$). 
Regarding the transition between these zones, we explicitly compared the relevant energy loss timescales. 
While the volume expansion implies adiabatic cooling, this channel is negligible for the high-energy particle population. 

The external photon field is assumed constant along the jet, originating from isotropic scattering of accretion-disk photons by the MHD wind (Section~\ref{sec:2}). The magnetic field decreases with distance as $B \propto 1/z$, consistent with magnetic flux conservation in a conical jet.

\begin{figure}[htbp]
    \centering
    \includegraphics[width=0.6\linewidth]{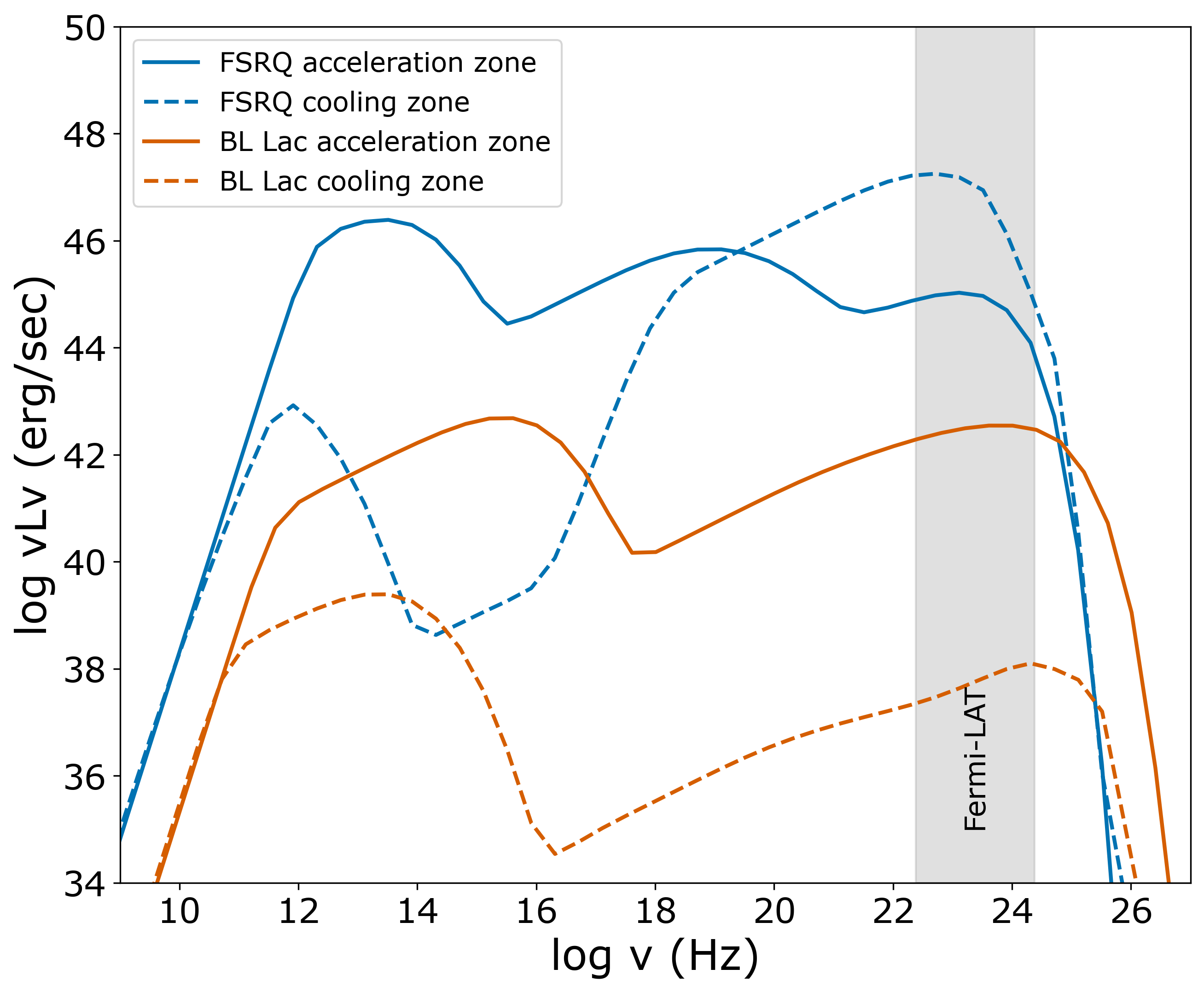}
    \caption{Results for FSRQ and BL Lac objects according to the two-zone model. Straight lines represent the emission from the acceleration zone, and dotted lines represent the emission from the cooling zone. The shaded region depicts the Fermi $\gamma$-ray energy band.}
    \label{fig:sed_2}
\end{figure}

In the two-zone framework, as it is depicted in Fig..\ref{fig:sed_2}.  BL Lac and FSRQ objects exhibit distinct emission patterns, reflecting their different mass accretion rates and the resulting variations in key physical parameters such as magnetic field, external photon energy density, and acceleration efficiency (see Table~\ref{tab:blazar_seq_input}). In BL Lac objects, the magnetic energy density dominates over the external photon field in both the acceleration and cooling zones. In the acceleration zone, synchrotron emission is therefore the primary radiative process. In contrast, in the cooling zone, although the magnetic field decreases with distance, it still satisfies $U_{B_{II}} \gg U_{\rm ext}$. As a result, the cooling zone contributes little to the total flux, and the overall spectrum is primarily determined by the acceleration zone (Fig.~\ref{fig:sed_2}). FSRQs, in contrast, display a more complex behavior. Synchrotron emission dominates in the acceleration zone, where $U_{B_I} > U_{\rm ext}$, but in the cooling zone the magnetic field decreases. In contrast, the external photon field remains approximately constant, \cite{BKM19}, leading to $U_{\rm ext} \gg U_{B_{II}}$. Consequently, external Compton scattering dominates the energy losses in the cooling zone, giving rise to the characteristic two-component SED observed in FSRQs, with low-energy synchrotron emission from the acceleration zone and high-energy external Compton emission from the cooling zone.

\begin{figure}[htbp]
    \centering
    \includegraphics[width=0.6\linewidth]{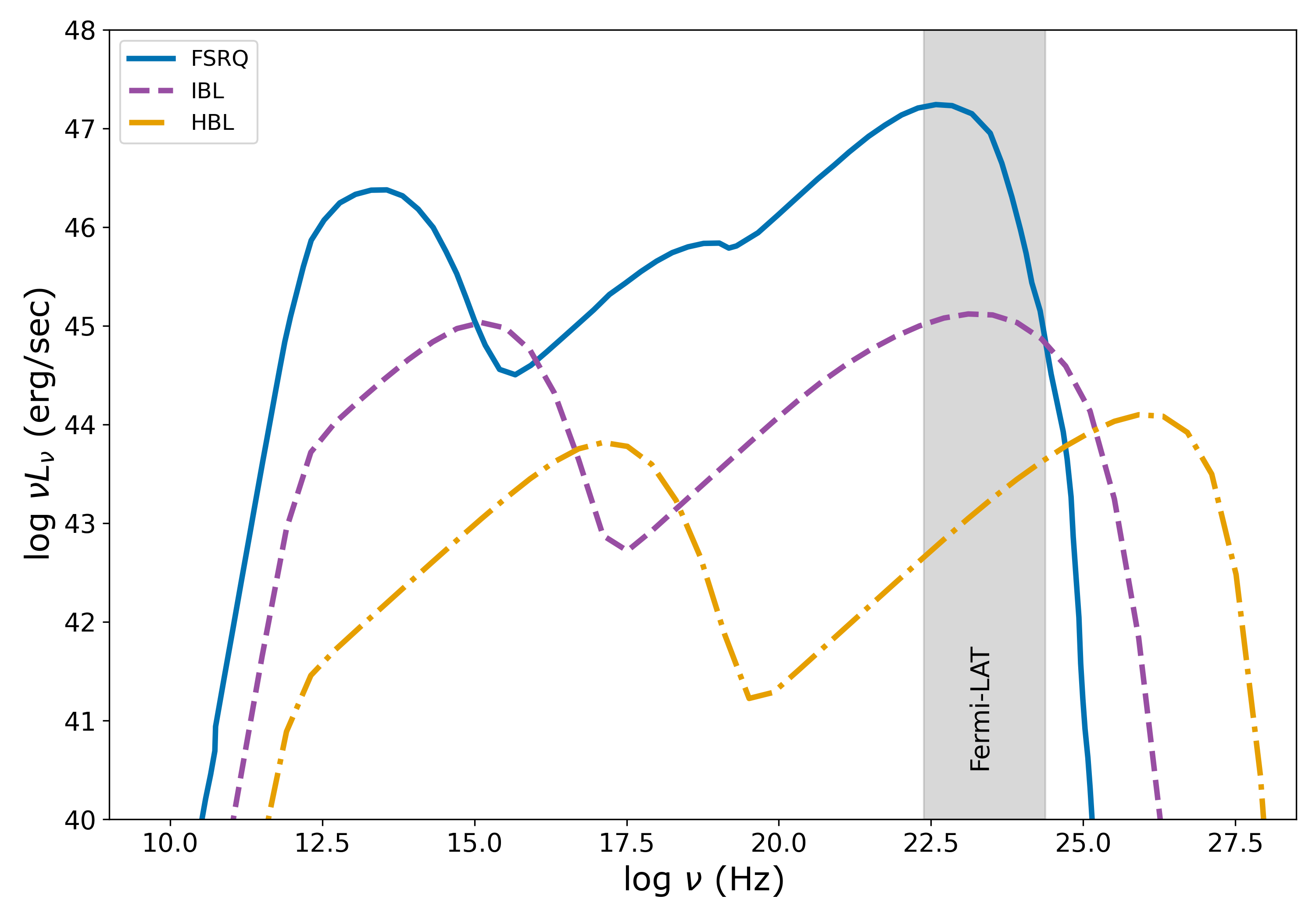}
    \caption{Theoretical Blazar Sequence according to the superposition of the two-zone emission by varying only the mass accretion rate, see Table~\ref{tab:blazar_seq_input} for the values of the input parameters. }
    \label{fig:sed_2zone}
\end{figure}
In Fig.~\ref{fig:sed_2zone} we present our results for a theoretical Blazar Sequence in the two-zone model (see Table~\ref{tab:blazar_seq_input} for the input parameters and Table~\ref{tab:blazar_seq_output} resulting spectral peak properties). In all classes, particles escape from the acceleration zone. FSRQs exhibit a strong $\gamma$-ray signature because electrons injected into the cooling zone interact with a dense external photon field, whereas in BL Lacs the escaping particles interact only weakly with the external field, and the cooling zone contributes less to the total flux. This two-zone approach is fully consistent with the statistical trends observed by \emph{Fermi}-LAT, including the Compton dominance, the $\gamma$-ray spectral indices, and the positions of the synchrotron and inverse Compton peaks. The values in Tables~\ref{tab:blazar_seq_input} and \ref{tab:blazar_seq_output} show that these trends can be reproduced by varying only the mass accretion rate, which emerges as the key physical driver of the sequence, as demonstrated in \cite{BKM19}. The same study shows that the black hole mass cannot account for the observational spread of the data: although both the black hole mass and the location of the emission region can affect details, their influence is subdominant compared to the role of $\dot{m}$. Furthermore, by allowing variations in the acceleration timescale, the electron injection properties, and the magnetic field efficiency, the two-zone framework can reproduce not only the systematic sequence but also the observed spread in the data, consistent with our previous one-zone study, \cite{BKM19,20BMK}. This demonstrates that the combination of mass accretion rate scaling and a two-zone treatment of particle escape provides a robust framework for capturing both the systematic and statistical features of the blazar population.

\begin{figure}[!htbp]
    \centering
    \includegraphics[width=\linewidth]{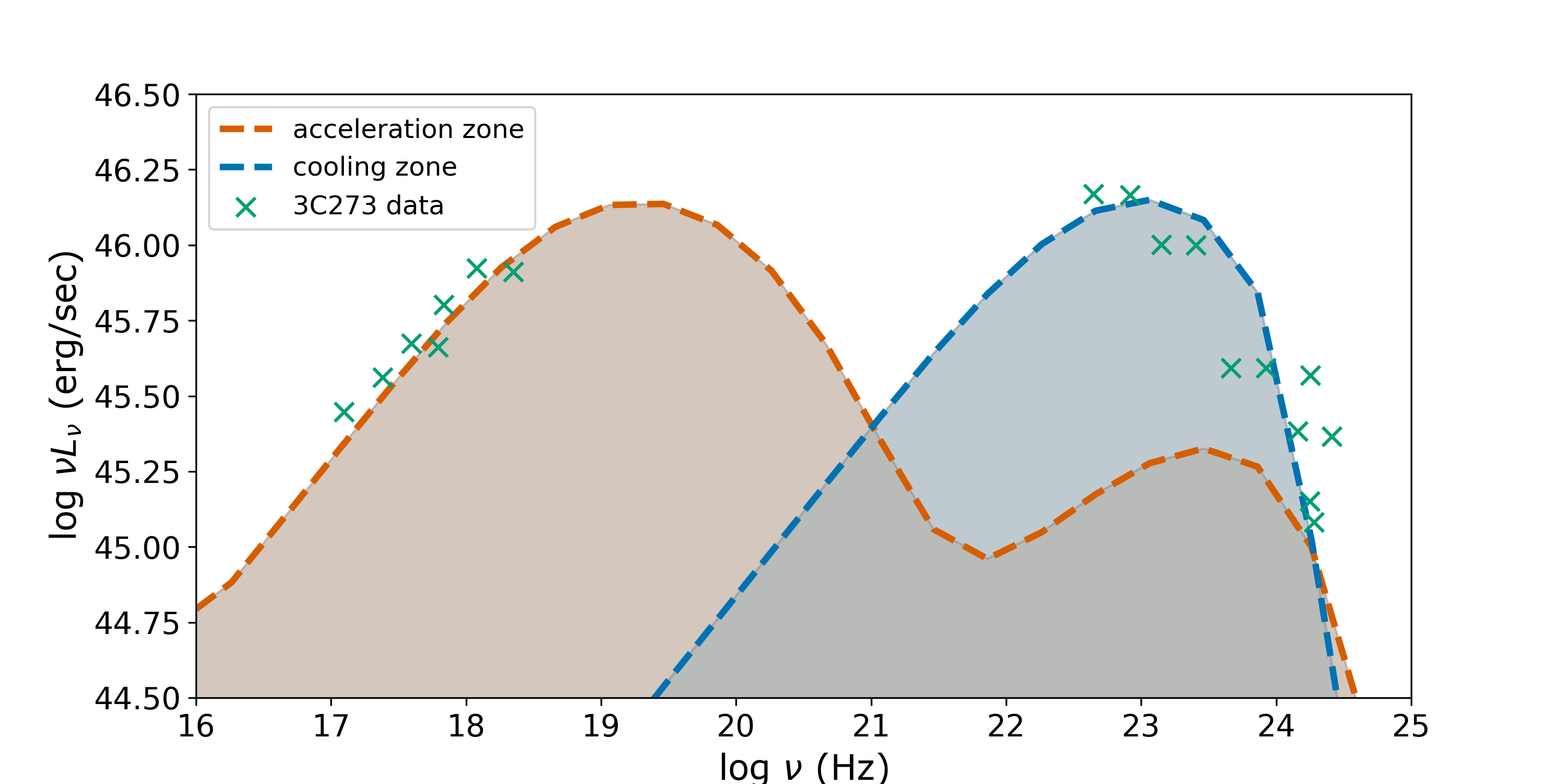}
    \caption{Application of the two-zone model to the FSRQ 3C273, focusing on the high-energy spectral range. The acceleration zone is located at $z=0.01$ pc, with a magnetic field strength of $B=1~\mathrm{G}$ and radius $R_I=5 \times 10^{15}$ cm, the cooling zone is one order of magnitude more distant and larger than the acceleration zone. The external photon field arises from isotropic scattering of disk photons on wind particles between radii $R_1 = 9 \times 10^{14}$ cm and $R_2 = 3 \times 10^{18}$ cm, with energy density $U_{\rm ext} = 2.5 \times 10^{-3}~\mathrm{erg\,cm^{-3}}$. The bulk Lorentz factor is $\Gamma=30$, the Doppler factor is $\delta=15$, and the characteristic disk temperature is $T_{\rm disk} = 3 \times 10^4$ K. Observational data are taken from \cite{GIommi12P}.}
    \label{fig:3c273}
\end{figure}

Figure~\ref{fig:3c273} illustrates the application of the two-zone model to the FSRQ 3C273, focusing on the X-ray and $\gamma$-ray bands. SSC primarily produces X-ray emission in the compact acceleration zone, where relativistic electrons upscatter synchrotron photons. The cooling zone dominates the $\gamma$-ray emission through external Compton scattering, where electrons interact with ambient photons from the MHD wind or the broad-line region. This two-zone scenario naturally explains the distinct emission components and spectral features observed in 3C273.
For the parameters adopted (e.g., $B_{II} \approx 0.1$\,G, $R_{II} \approx 3\times10^{17}$\,cm, and $u_{\mathrm{exp}} \approx 0.01 c$), the synchrotron cooling timescale $t_{\mathrm{syn}}$ is significantly shorter than the adiabatic expansion timescale 
$t_{\mathrm{ad}} \approx R/u_{\mathrm{exp}}$ for electrons with $\gamma \gtrsim 200$. 
Consequently, the radiating electrons lose their energy via synchrotron emission before they undergo significant adiabatic decompression.

\section{Conclusion}\label{sec:4}
In this work, we have successfully reproduced the theoretical Blazar Sequence by extending the model of \cite{BKM19}, using the mass accretion rate as the primary physical parameter to account for the broad diversity observed in blazar phenomenology. This demonstrates that a simple, physically grounded approach—where changes in accretion rate alone drive the differences between Flat Spectrum Radio Quasars (FSRQs) and BL Lacertae objects (BL Lacs)—is consistent with the blazar unification paradigm.

Our model solves the coupled kinetic equations for electrons and photons self-consistently, including Fermi acceleration, synchrotron, and inverse Compton losses, within a two-zone emission framework. Electrons accelerate in a compact region and partially escape into a larger cooling zone, naturally producing the distinct spectral components characteristic of blazars. The competition between magnetic and external photon field energy densities, modulated by the accretion rate, explains the dominance of synchrotron and SSC emission in BL Lacs versus strong external Compton emission in FSRQs.

Importantly, this approach reproduces not only the main trends of the Blazar Sequence but also the observed statistical distributions reported by Fermi-LAT, including Compton Dominance, $\gamma$-ray spectral indices, and the locations of the synchrotron and inverse Compton peaks. Application of the model to the well-studied FSRQ 3C273 captures key features of its broadband SED, confirming that variations in accretion power are sufficient to account for its high-energy emission.

An interesting implication of the two-zone framework is that, in a flaring state, FSRQs may exhibit a time delay in the $\gamma$-ray emission relative to the X-rays. Initially, $\gamma$-rays are produced simultaneously with X-rays in the acceleration zone, but additional $\gamma$-ray emission arises from the cooling zone as electrons propagate and radiate. The duration of the $\gamma$-ray pulse is therefore set primarily by the electron cooling timescale, since freshly accelerated ones do not immediately replace cooled electrons. Detailed modeling of this effect requires time-dependent approach. A similar argument applies to the radio emission, which is expected to appear later due to the longer cooling and propagation timescales of the lower-energy electrons.

Overall, our results underscore the central role of the mass accretion rate in shaping blazar spectra within a unified theoretical framework grounded in particle acceleration physics and radiative transfer. The inclusion of particle escape and spatially structured emission zones provides a more realistic interpretation of blazar spectra. Future refinements, including jet stratification, time variability, and polarization studies, will further enhance the predictive power of the model. However, the present work establishes a robust foundation that links accretion physics, jet emission and blazar phenomenology.

\begin{acknowledgements}
We thank the anonymous referee for their useful comments. We thank Yannis Liodakis for the helpful feedback on the manuscript and Fabrizio Tavecchio for valuable discussions. We have used the following Python libraries: Numpy \citep{numpy}, Matplotlib \citep{matplotlib}, Scipy \citep{scipy}
\end{acknowledgements}

\bibliographystyle{aa}
\bibliography{bib} 

@INPROCEEDINGS{20BMK,
       author = {{Boula}, S. and {Kazanas}, D. and {Mastichiadis}, A.},
        title = "{Mhd Accretion Disk Winds And The Blazar Sequence}",
    booktitle = {High Energy Phenomena in Relativistic Outflows VII},
         year = 2019,
        month = jul,
          eid = {9},
        pages = {9},
       adsurl = {https://ui.adsabs.harvard.edu/abs/2019hepr.confE...9B}
}

@ARTICLE{Blandford19,
       author = {{Blandford}, Roger and {Meier}, David and {Readhead}, Anthony},
        title = "{Relativistic Jets from Active Galactic Nuclei}",
      journal = {\araa},
         year = 2019,
        month = aug,
       volume = {57},
        pages = {467-509},
          doi = {10.1146/annurev-astro-081817-051948},
archivePrefix = {arXiv},
       eprint = {1812.06025},
 primaryClass = {astro-ph.HE},
       adsurl = {https://ui.adsabs.harvard.edu/abs/2019ARA&A..57..467B}
}

@ARTICLE{BM22,
       author = {{Boula}, S. and {Mastichiadis}, A.},
        title = "{Expanding one-zone model for blazar emission}",
      journal = {\aap},
         year = 2022,
        month = jan,
       volume = {657},
          eid = {A20},
        pages = {A20},
          doi = {10.1051/0004-6361/202142126},
archivePrefix = {arXiv},
       eprint = {2110.05325},
 primaryClass = {astro-ph.HE},
       adsurl = {https://ui.adsabs.harvard.edu/abs/2022A&A...657A..20B}
}

@ARTICLE{scipy,
  author  = {Virtanen, Pauli and Gommers, Ralf and Oliphant, Travis E. and
            Haberland, Matt and Reddy, Tyler and Cournapeau, David and
            Burovski, Evgeni and Peterson, Pearu and Weckesser, Warren and
            Bright, Jonathan and {van der Walt}, St{\'e}fan J. and
            Brett, Matthew and Wilson, Joshua and Millman, K. Jarrod and
            Mayorov, Nikolay and Nelson, Andrew R. J. and Jones, Eric and
            Kern, Robert and Larson, Eric and Carey, C J and
            Polat, {\.I}lhan and Feng, Yu and Moore, Eric W. and
            {VanderPlas}, Jake and Laxalde, Denis and Perktold, Josef and
            Cimrman, Robert and Henriksen, Ian and Quintero, E. A. and
            Harris, Charles R. and Archibald, Anne M. and
            Ribeiro, Ant{\^o}nio H. and Pedregosa, Fabian and
            {van Mulbregt}, Paul and {SciPy 1.0 Contributors}},
  title   = {{{SciPy} 1.0: Fundamental Algorithms for Scientific
            Computing in Python}},
  journal = {Nature Methods},
  year    = {2020},
  volume  = {17},
  pages   = {261--272},
  adsurl  = {https://rdcu.be/b08Wh},
  doi     = {10.1038/s41592-019-0686-2},
}

@Article{numpy,
 title         = {Array programming with {NumPy}},
 author        = {Charles R. Harris and K. Jarrod Millman and St{\'{e}}fan J.
                 van der Walt and Ralf Gommers and Pauli Virtanen and David
                 Cournapeau and Eric Wieser and Julian Taylor and Sebastian
                 Berg and Nathaniel J. Smith and Robert Kern and Matti Picus
                 and Stephan Hoyer and Marten H. van Kerkwijk and Matthew
                 Brett and Allan Haldane and Jaime Fern{\'{a}}ndez del
                 R{\'{i}}o and Mark Wiebe and Pearu Peterson and Pierre
                 G{\'{e}}rard-Marchant and Kevin Sheppard and Tyler Reddy and
                 Warren Weckesser and Hameer Abbasi and Christoph Gohlke and
                 Travis E. Oliphant},
 year          = {2020},
 month         = sep,
 journal       = {Nature},
 volume        = {585},
 number        = {7825},
 pages         = {357--362},
 doi           = {10.1038/s41586-020-2649-2},
 publisher     = {Springer Science and Business Media {LLC}},
 url           = {https://doi.org/10.1038/s41586-020-2649-2}
}

@ARTICLE{tavecchio98,
  author = {Tavecchio, F. and Maraschi, L. and Ghisellini, G.},
  title = {Constraints on the Physical Parameters of TeV Blazars},
  journal = {Astrophysical Journal},
  year = {1998},
  volume = {509},
  pages = {608--619},
  doi = {10.1086/306526},
}

@ARTICLE{dermer92,
       author = {{Dermer}, C.~D. and {Schlickeiser}, R. and {Mastichiadis}, A.},
        title = "{High-energy gamma radiation from extragalactic radio sources.}",
      journal = {\aap},
         year = 1992,
        month = mar,
       volume = {256},
        pages = {L27-L30},
       adsurl = {https://ui.adsabs.harvard.edu/abs/1992A&A...256L..27D}
}

@ARTICLE{sikora94,
       author = {{Sikora}, Marek and {Begelman}, Mitchell C. and {Rees}, Martin J.},
        title = "{Comptonization of Diffuse Ambient Radiation by a Relativistic Jet: The Source of Gamma Rays from Blazars?}",
      journal = {\apj},
         year = 1994,
        month = jan,
       volume = {421},
        pages = {153},
          doi = {10.1086/173633},
       adsurl = {https://ui.adsabs.harvard.edu/abs/1994ApJ...421..153S}
}

@ARTICLE{tavecchio21,
       author = {{Tavecchio}, F.},
        title = "{Constraining the shear acceleration model for the X-ray emission of large-scale extragalactic jets}",
      journal = {\mnras},
         year = 2021,
        month = mar,
       volume = {501},
       number = {4},
        pages = {6199-6207},
          doi = {10.1093/mnras/staa4009},
archivePrefix = {arXiv},
       eprint = {2011.03264},
 primaryClass = {astro-ph.HE},
       adsurl = {https://ui.adsabs.harvard.edu/abs/2021MNRAS.501.6199T}
}

@ARTICLE{Liodakis2022,
       author = {{Liodakis}, Ioannis and {Marscher}, Alan P. and {Agudo}, Iv{\'a}n and {Berdyugin}, Andrei V. and {Bernardos}, Maria I. and {Bonnoli}, Giacomo and {Borman}, George A. and {Casadio}, Carolina and {Casanova}, V{\'\i}ctor and {Cavazzuti}, Elisabetta and {Rodriguez Cavero}, Nicole and {Di Gesu}, Laura and {Di Lalla}, Niccol{\'o} and {Donnarumma}, Immacolata and {Ehlert}, Steven R. and {Errando}, Manel and {Escudero}, Juan and {Garc{\'\i}a-Comas}, Maya and {Ag{\'\i}s-Gonz{\'a}lez}, Beatriz and {Husillos}, C{\'e}sar and {Jormanainen}, Jenni and {Jorstad}, Svetlana G. and {Kagitani}, Masato and {Kopatskaya}, Evgenia N. and {Kravtsov}, Vadim and {Krawczynski}, Henric and {Lindfors}, Elina and {Larionova}, Elena G. and {Madejski}, Grzegorz M. and {Marin}, Fr{\'e}d{\'e}ric and {Marchini}, Alessandro and {Marshall}, Herman L. and {Morozova}, Daria A. and {Massaro}, Francesco and {Masiero}, Joseph R. and {Mawet}, Dimitri and {Middei}, Riccardo and {Millar-Blanchaer}, Maxwell A. and {Myserlis}, Ioannis and {Negro}, Michela and {Nilsson}, Kari and {O'Dell}, Stephen L. and {Omodei}, Nicola and {Pacciani}, Luigi and {Paggi}, Alessandro and {Panopoulou}, Georgia V. and {Peirson}, Abel L. and {Perri}, Matteo and {Petrucci}, Pierre-Olivier and {Poutanen}, Juri and {Puccetti}, Simonetta and {Romani}, Roger W. and {Sakanoi}, Takeshi and {Savchenko}, Sergey S. and {Sota}, Alfredo and {Tavecchio}, Fabrizio and {Tinyanont}, Samaporn and {Vasilyev}, Andrey A. and {Weaver}, Zachary R. and {Zhovtan}, Alexey V. and {Antonelli}, Lucio A. and {Bachetti}, Matteo and {Baldini}, Luca and {Baumgartner}, Wayne H. and {Bellazzini}, Ronaldo and {Bianchi}, Stefano and {Bongiorno}, Stephen D. and {Bonino}, Raffaella and {Brez}, Alessandro and {Bucciantini}, Niccol{\'o} and {Capitanio}, Fiamma and {Castellano}, Simone and {Ciprini}, Stefano and {Costa}, Enrico and {De Rosa}, Alessandra and {Del Monte}, Ettore and {Di Marco}, Alessandro and {Doroshenko}, Victor and {Dov{\v{c}}iak}, Michal and {Enoto}, Teruaki and {Evangelista}, Yuri and {Fabiani}, Sergio and {Ferrazzoli}, Riccardo and {Garcia}, Javier A. and {Gunji}, Shuichi and {Hayashida}, Kiyoshi and {Heyl}, Jeremy and {Iwakiri}, Wataru and {Karas}, Vladimir and {Kitaguchi}, Takao and {Kolodziejczak}, Jeffery J. and {La Monaca}, Fabio and {Latronico}, Luca and {Maldera}, Simone and {Manfreda}, Alberto and {Marinucci}, Andrea and {Matt}, Giorgio and {Mitsuishi}, Ikuyuki and {Mizuno}, Tsunefumi and {Muleri}, Fabio and {Ng}, Stephen C. -Y. and {Oppedisano}, Chiara and {Papitto}, Alessandro and {Pavlov}, George G. and {Pesce-Rollins}, Melissa and {Pilia}, Maura and {Possenti}, Andrea and {Ramsey}, Brian D. and {Rankin}, John and {Ratheesh}, Ajay and {Sgr{\'o}}, Carmelo and {Slane}, Patrick and {Soffitta}, Paolo and {Spandre}, Gloria and {Tamagawa}, Toru and {Taverna}, Roberto and {Tawara}, Yuzuru and {Tennant}, Allyn F. and {Thomas}, Nicolas E. and {Tombesi}, Francesco and {Trois}, Alessio and {Tsygankov}, Sergey and {Turolla}, Roberto and {Vink}, Jacco and {Weisskopf}, Martin C. and {Wu}, Kinwah and {Xie}, Fei and {Zane}, Silvia},
        title = "{Polarized blazar X-rays imply particle acceleration in shocks}",
      journal = {\nat},
         year = 2022,
        month = nov,
       volume = {611},
       number = {7937},
        pages = {677-681},
          doi = {10.1038/s41586-022-05338-0},
archivePrefix = {arXiv},
       eprint = {2209.06227},
 primaryClass = {astro-ph.HE},
       adsurl = {https://ui.adsabs.harvard.edu/abs/2022Natur.611..677L}
}

@Article{matplotlib,
  Author    = {Hunter, J. D.},
  Title     = {Matplotlib: A 2D graphics environment},
  Journal   = {Computing in Science \& Engineering},
  Volume    = {9},
  Number    = {3},
  Pages     = {90--95},
  publisher = {IEEE COMPUTER SOC},
  doi       = {10.1109/MCSE.2007.55},
  year      = 2007
}

@ARTICLE{Giommi12P,
       author = {{Giommi}, P. and {Polenta}, G. and {L{\"a}hteenm{\"a}ki}, A. and {Thompson}, D.~J. and {Capalbi}, M. and {Cutini}, S. and {Gasparrini}, D. and {Gonz{\'a}lez-Nuevo}, J. and {Le{\'o}n-Tavares}, J. and {L{\'o}pez-Caniego}, M. and {Mazziotta}, M.~N. and {Monte}, C. and {Perri}, M. and {Rain{\`o}}, S. and {Tosti}, G. and {Tramacere}, A. and {Verrecchia}, F. and {Aller}, H.~D. and {Aller}, M.~F. and {Angelakis}, E. and {Bastieri}, D. and {Berdyugin}, A. and {Bonaldi}, A. and {Bonavera}, L. and {Burigana}, C. and {Burrows}, D.~N. and {Buson}, S. and {Cavazzuti}, E. and {Chincarini}, G. and {Colafrancesco}, S. and {Costamante}, L. and {Cuttaia}, F. and {D'Ammando}, F. and {de Zotti}, G. and {Frailis}, M. and {Fuhrmann}, L. and {Galeotta}, S. and {Gargano}, F. and {Gehrels}, N. and {Giglietto}, N. and {Giordano}, F. and {Giroletti}, M. and {Keih{\"a}nen}, E. and {King}, O. and {Krichbaum}, T.~P. and {Lasenby}, A. and {Lavonen}, N. and {Lawrence}, C.~R. and {Leto}, C. and {Lindfors}, E. and {Mandolesi}, N. and {Massardi}, M. and {Max-Moerbeck}, W. and {Michelson}, P.~F. and {Mingaliev}, M. and {Natoli}, P. and {Nestoras}, I. and {Nieppola}, E. and {Nilsson}, K. and {Partridge}, B. and {Pavlidou}, V. and {Pearson}, T.~J. and {Procopio}, P. and {Rachen}, J.~P. and {Readhead}, A. and {Reeves}, R. and {Reimer}, A. and {Reinthal}, R. and {Ricciardi}, S. and {Richards}, J. and {Riquelme}, D. and {Saarinen}, J. and {Sajina}, A. and {Sandri}, M. and {Savolainen}, P. and {Sievers}, A. and {Sillanp{\"a}{\"a}}, A. and {Sotnikova}, Y. and {Stevenson}, M. and {Tagliaferri}, G. and {Takalo}, L. and {Tammi}, J. and {Tavagnacco}, D. and {Terenzi}, L. and {Toffolatti}, L. and {Tornikoski}, M. and {Trigilio}, C. and {Turunen}, M. and {Umana}, G. and {Ungerechts}, H. and {Villa}, F. and {Wu}, J. and {Zacchei}, A. and {Zensus}, J.~A. and {Zhou}, X.},
        title = "{Simultaneous Planck, Swift, and Fermi observations of X-ray and {\ensuremath{\gamma}}-ray selected blazars}",
      journal = {\aap},
         year = 2012,
        month = may,
       volume = {541},
          eid = {A160},
        pages = {A160},
          doi = {10.1051/0004-6361/201117825},
archivePrefix = {arXiv},
       eprint = {1108.1114},
 primaryClass = {astro-ph.CO},
       adsurl = {https://ui.adsabs.harvard.edu/abs/2012A&A...541A.160G}
}

@ARTICLE{KRM98,
       author = {{Kirk}, J.~G. and {Rieger}, F.~M. and {Mastichiadis}, A.},
        title = "{Particle acceleration and synchrotron emission in blazar jets}",
      journal = {\aap},
         year = 1998,
        month = may,
       volume = {333},
        pages = {452-458},
archivePrefix = {arXiv},
       eprint = {astro-ph/9801265},
 primaryClass = {astro-ph},
       adsurl = {https://ui.adsabs.harvard.edu/abs/1998A&A...333..452K}
}

@ARTICLE{AntonMill,
   author = {{Antonucci}, R.~R.~J. and {Miller}, J.~S.},
    title = "{Spectropolarimetry and the nature of NGC 1068}",
  journal = {\apj},
     year = 1985,
    month = oct,
   volume = 297,
    pages = {621-632},
      doi = {10.1086/163559},
   adsurl = {http://adsabs.harvard.edu/abs/1985ApJ...297..621A}
}

@ARTICLE{Behar09,
   author = {{Behar}, E.},
    title = "{Density Profiles in Seyfert Outflows}",
  journal = {\apj},
archivePrefix = "arXiv",
   eprint = {0908.0539},
     year = 2009,
    month = oct,
   volume = 703,
    pages = {1346-1351},
      doi = {10.1088/0004-637X/703/2/1346},
   adsurl = {http://adsabs.harvard.edu/abs/2009ApJ...703.1346B}
}

@ARTICLE{CL94,
       author = {{Contopoulos}, J. and {Lovelace}, R.~V.~E.},
        title = "{Magnetically Driven Jets and Winds: Exact Solutions}",
      journal = {\apj},
         year = "1994",
        month = "Jul",
       volume = {429},
        pages = {139},
          doi = {10.1086/174307},
       adsurl = {https://ui.adsabs.harvard.edu/abs/1994ApJ...429..139C}
}

@ARTICLE{Finke13,
   author = {{Finke}, J.~D.},
    title = "{Compton Dominance and the Blazar Sequence}",
  journal = {\apj},
archivePrefix = "arXiv",
   eprint = {1212.0869},
 primaryClass = "astro-ph.HE",
     year = 2013,
    month = feb,
   volume = 763,
      eid = {134},
    pages = {134},
      doi = {10.1088/0004-637X/763/2/134},
   adsurl = {http://adsabs.harvard.edu/abs/2013ApJ...763..134F}
}

@ARTICLE{BKM19,
       author = {{Boula}, Stella and {Kazanas}, Demosthenes and {Mastichiadis}, Apostolos},
        title = "{Accretion disc MHD winds and blazar classification}",
      journal = {\mnras},
         year = "2019",
        month = "Jan",
       volume = {482},
       number = {1},
        pages = {L80-L84},
          doi = {10.1093/mnrasl/sly189},
archivePrefix = {arXiv},
       eprint = {1810.01796},
 primaryClass = {astro-ph.HE},
       adsurl = {https://ui.adsabs.harvard.edu/abs/2019MNRAS.482L..80B}
}

@ARTICLE{Foss98,
   author = {{Fossati}, G. and {Maraschi}, L. and {Celotti}, A. and {Comastri}, A. and 
	{Ghisellini}, G.},
    title = "{A unifying view of the spectral energy distributions of blazars}",
  journal = {\mnras},
   eprint = {astro-ph/9804103},
     year = 1998,
    month = sep,
   volume = 299,
    pages = {433-448},
      doi = {10.1046/j.1365-8711.1998.01828.x},
   adsurl = {http://adsabs.harvard.edu/abs/1998MNRAS.299..433F}
}

@ARTICLE{FKCB,
   author = {{Fukumura}, K. and {Kazanas}, D. and {Contopoulos}, I. and {Behar}, E.
	},
    title = "{Magnetohydrodynamic Accretion Disk Winds as X-ray Absorbers in Active Galactic Nuclei}",
  journal = {\apj},
archivePrefix = "arXiv",
   eprint = {0910.3001},
 primaryClass = "astro-ph.HE",
     year = 2010,
    month = may,
   volume = 715,
    pages = {636-650},
      doi = {10.1088/0004-637X/715/1/636},
   adsurl = {http://adsabs.harvard.edu/abs/2010ApJ...715..636F}
}

@ARTICLE{FKCB17,
   author = {{Fukumura}, K. and {Kazanas}, D. and {Shrader}, C. and {Behar}, E. and 
	{Tombesi}, F. and {Contopoulos}, I.},
    title = "{Magnetic origin of black hole winds across the mass scale}",
  journal = {Nature Astronomy},
archivePrefix = "arXiv",
   eprint = {1702.02197},
 primaryClass = "astro-ph.HE",
     year = 2017,
    month = mar,
   volume = 1,
      eid = {0062},
    pages = {0062},
      doi = {10.1038/s41550-017-0062},
   adsurl = {http://adsabs.harvard.edu/abs/2017NatAs...1E..62F}
}

@ARTICLE{Ghis17,
   author = {{Ghisellini}, G. and {Righi}, C. and {Costamante}, L. and {Tavecchio}, F.
	},
    title = "{The Fermi blazar sequence}",
  journal = {\mnras},
archivePrefix = "arXiv",
   eprint = {1702.02571},
 primaryClass = "astro-ph.HE",
     year = 2017,
    month = jul,
   volume = 469,
    pages = {255-266},
      doi = {10.1093/mnras/stx806},
   adsurl = {http://adsabs.harvard.edu/abs/2017MNRAS.469..255G}
}

@ARTICLE{KK94,
   author = {{Konigl}, A. and {Kartje}, J.~F.},
    title = "{Disk-driven hydromagnetic winds as a key ingredient of active galactic nuclei unification schemes}",
  journal = {\apj},
     year = 1994,
    month = oct,
   volume = 434,
    pages = {446-467},
      doi = {10.1086/174746},
   adsurl = {http://adsabs.harvard.edu/abs/1994ApJ...434..446K}
}

@ARTICLE{Marscher10,
   author = {{Marscher}, A.~P. and {Jorstad}, S.~G. and {Larionov}, V.~M. and 
	{Aller}, M.~F. and {Aller}, H.~D. and {L{\"a}hteenm{\"a}ki}, A. and 
	{Agudo}, I. and {Smith}, P.~S. and {Gurwell}, M. and {Hagen-Thorn}, V.~A. and 
	{Konstantinova}, T.~S. and {Larionova}, E.~G. and {Larionova}, L.~V. and 
	{Melnichuk}, D.~A. and {Blinov}, D.~A. and {Kopatskaya}, E.~N. and 
	{Troitsky}, I.~S. and {Tornikoski}, M. and {Hovatta}, T. and 
	{Schmidt}, G.~D. and {D'Arcangelo}, F.~D. and {Bhattarai}, D. and 
	{Taylor}, B. and {Olmstead}, A.~R. and {Manne-Nicholas}, E. and 
	{Roca-Sogorb}, M. and {G{\'o}mez}, J.~L. and {McHardy}, I.~M. and 
	{Kurtanidze}, O. and {Nikolashvili}, M.~G. and {Kimeridze}, G.~N. and 
	{Sigua}, L.~A.},
    title = "{Probing the Inner Jet of the Quasar PKS 1510-089 with Multi-Waveband Monitoring During Strong Gamma-Ray Activity}",
  journal = {\apjl},
archivePrefix = "arXiv",
   eprint = {1001.2574},
 primaryClass = "astro-ph.CO",
     year = 2010,
    month = feb,
   volume = 710,
    pages = {L126-L131},
      doi = {10.1088/2041-8205/710/2/L126},
   adsurl = {http://adsabs.harvard.edu/abs/2010ApJ...710L.126M}
}

@ARTICLE{4th,
       author = {{The Fermi-LAT collaboration}},
        title = "{The Fourth Catalog of Active Galactic Nuclei Detected by the Fermi Large Area Telescope}",
      journal = {arXiv e-prints},
         year = "2019",
        month = "May",
          eid = {arXiv:1905.10771},
        pages = {arXiv:1905.10771},
archivePrefix = {arXiv},
       eprint = {1905.10771},
 primaryClass = {astro-ph.HE},
       adsurl = {https://ui.adsabs.harvard.edu/abs/2019arXiv190510771T}
}

@ARTICLE{MK95,
       author = {{Mastichiadis}, A. and {Kirk}, J.~G.},
        title = "{Self-consistent particle acceleration in active galactic nuclei.}",
      journal = {\aap},
         year = "1995",
        month = "Mar",
       volume = {295},
        pages = {613},
       adsurl = {https://ui.adsabs.harvard.edu/abs/1995A&A...295..613M}
}

@ARTICLE{RiegerMannheim2002,
       author = {{Rieger}, F.~M. and {Mannheim}, K.},
        title = "{Particle acceleration in rotating and shearing jets from AGN}",
      journal = {\aap},
         year = 2002,
        month = dec,
       volume = {396},
        pages = {833-846},
          doi = {10.1051/0004-6361:20021457},
archivePrefix = {arXiv},
       eprint = {astro-ph/0210286},
 primaryClass = {astro-ph},
       adsurl = {https://ui.adsabs.harvard.edu/abs/2002A&A...396..833R}
}

@ARTICLE{Blandford1978,
       author = {{Blandford}, R.~D. and {Ostriker}, J.~P.},
        title = "{Particle acceleration by astrophysical shocks.}",
      journal = {\apjl},
         year = 1978,
        month = apr,
       volume = {221},
        pages = {L29-L32},
          doi = {10.1086/182658},
       adsurl = {https://ui.adsabs.harvard.edu/abs/1978ApJ...221L..29B}
}

@ARTICLE{Kirk2000,
       author = {{Kirk}, J.~G. and {Duffy}, P.},
        title = "{TOPICAL REVIEW: Particle acceleration and relativistic shocks}",
      journal = {Journal of Physics G Nuclear Physics},
         year = 1999,
        month = aug,
       volume = {25},
       number = {8},
        pages = {R163-R194},
          doi = {10.1088/0954-3899/25/8/201},
archivePrefix = {arXiv},
       eprint = {astro-ph/9905069},
 primaryClass = {astro-ph},
       adsurl = {https://ui.adsabs.harvard.edu/abs/1999JPhG...25R.163K}
}

@ARTICLE{M2020,
       author = {{Matthews}, James H. and {Bell}, Anthony R. and {Blundell}, Katherine M.},
        title = "{Particle acceleration in astrophysical jets}",
      journal = {\nar},
         year = 2020,
        month = sep,
       volume = {89},
          eid = {101543},
        pages = {101543},
          doi = {10.1016/j.newar.2020.101543},
archivePrefix = {arXiv},
       eprint = {2003.06587},
 primaryClass = {astro-ph.HE},
       adsurl = {https://ui.adsabs.harvard.edu/abs/2020NewAR..8901543M}
}

@ARTICLE{2006KGMT,
       author = {{Katarzy{\'n}ski}, K. and {Ghisellini}, G. and {Mastichiadis}, A. and {Tavecchio}, F. and {Maraschi}, L.},
        title = "{Stochastic particle acceleration and synchrotron self-Compton radiation in TeV blazars}",
      journal = {\aap},
         year = 2006,
        month = jul,
       volume = {453},
       number = {1},
        pages = {47-56},
          doi = {10.1051/0004-6361:20054176},
archivePrefix = {arXiv},
       eprint = {astro-ph/0603362},
 primaryClass = {astro-ph},
       adsurl = {https://ui.adsabs.harvard.edu/abs/2006A&A...453...47K}
}

@ARTICLE{rieger2019,
       author = {{Rieger}, Frank M.},
        title = "{An Introduction to Particle Acceleration in Shearing Flows}",
      journal = {Galaxies},
         year = 2019,
        month = sep,
       volume = {7},
       number = {3},
          eid = {78},
        pages = {78},
          doi = {10.3390/galaxies7030078},
archivePrefix = {arXiv},
       eprint = {1909.07237},
 primaryClass = {astro-ph.HE},
       adsurl = {https://ui.adsabs.harvard.edu/abs/2019Galax...7...78R}
}

@ARTICLE{sironi2025,
       author = {{Sironi}, Lorenzo and {Uzdensky}, Dmitri A. and {Giannios}, Dimitrios},
        title = "{Relativistic Magnetic Reconnection in Astrophysical Plasmas: A Powerful Mechanism of Nonthermal Emission}",
      journal = {arXiv e-prints},
         year = 2025,
        month = jun,
          eid = {arXiv:2506.02101},
        pages = {arXiv:2506.02101},
          doi = {10.48550/arXiv.2506.02101},
archivePrefix = {arXiv},
       eprint = {2506.02101},
 primaryClass = {astro-ph.HE},
       adsurl = {https://ui.adsabs.harvard.edu/abs/2025arXiv250602101S}
}

@misc{Shirin25,
      title={Spectral curvature and breaks from Fermi acceleration at oblique shocks}, 
      author={Asma {Shirin T} and Brian Reville and Nils W. Schween and Florian Schulze and John G. Kirk},
      year={2025},
      eprint={2511.01635},
      archivePrefix={arXiv},
      primaryClass={astro-ph.HE},
      url={https://arxiv.org/abs/2511.01635}, 
}

\begin{appendix}
\section{Parameters and results of the theoretical blazar sequence}
In Tables~\ref{tab:blazar_seq_input} and \ref{tab:blazar_seq_output} we present
the input parameters of the theoretical blazar sequence and the main
characteristics of the produced spectral energy distributions that are in
agreement with the observational trends of fermi blazars \citep{4th}.

\FloatBarrier

\begin{table}[!ht]
\caption{Input parameters (log scale) of the theoretical blazar sequence for
$\mathcal{M}=10^8$ (see Fig.~\ref{fig:sed_2zone}).}
\label{tab:blazar_seq_input}
\centering
\setlength{\tabcolsep}{6pt}
\renewcommand{\arraystretch}{1.2}
\footnotesize
\begin{tabular}{lcccc}
\hline
Class & $\dot{m}$ & $B$ (G) & $U_{\rm ext}$ (erg cm$^{-3}$) & $A_{\rm acc}$ \\
\hline
FSRQ & $-0.5$ & $1$  & $-2.6$ & $-4$ \\
LBL  & $-1.5$ & $0$  & $-5.6$ & $-5$ \\
HBL  & $-2.5$ & $-1$ & $-8.6$ & $-6$ \\
\hline
\end{tabular}
\end{table}

\begin{table}[!ht]
\caption{Resulting peak frequencies and luminosities of the synchrotron and
inverse Compton components in the theoretical blazar sequence.}
\label{tab:blazar_seq_output}
\centering
\setlength{\tabcolsep}{3.5pt}
\renewcommand{\arraystretch}{1.}
\footnotesize
\begin{tabular}{lcccccc}
\hline
Class & $\nu_{\rm syn}^{\rm pk}$ & $L_{\rm syn}^{\rm pk}$
& $\nu_{\rm IC}^{\rm pk}$ & $L_{\rm IC}^{\rm pk}$ & $\Gamma_\gamma$ & $\log(\mathrm{CD})$ \\
      & (log Hz) & (log erg s$^{-1}$)
      & (log Hz) & (log erg s$^{-1}$) & & \\
\hline
FSRQ & 13.2 & 46.4 & 22.8 & 47.3 & 3.00 & 0.92 \\
LBL  & 15.5 & 45.0 & 23.3 & 45.1 & 2.14 & 0.12 \\
HBL  & 17.1 & 43.8 & 26.3 & 44.0 & 1.46 & 0.21 \\
\hline
\end{tabular}

\tablefoot{The acceleration zone is at a distance of $z=0.01~\mathrm{pc}$ from
the central black hole, assumed to have a mass
$\mathcal{M} = 10^8\,M_{\odot}$. The external photon field is produced from
isotropic scattering of disk photons on wind particles between radii
$R_1 = 9 \times 10^{14}~\mathrm{cm}$ and
$R_2 = 3 \times 10^{18}~\mathrm{cm}$. The efficiencies of the magnetic field
and the external photon field are $\eta_{\rm b} = 0.01$ and $\epsilon = 0.05$,
respectively. The number of electrons injected in the acceleration process,
assumed to be of Type I Fermi, depends linearly on the mass accretion rate.
The bulk Lorentz factor is $\Gamma = 30$ and the Doppler factor is $\delta = 15$.
The characteristic temperature of the accretion disk is
$T_{\rm disk} = 3 \times 10^3~\mathrm{K}$.}
\end{table}

\end{appendix}
\end{document}